# On the equivocal nature of the mass absorption curves

Pawel Rochowski

Institute of Experimental Physics, Faculty of Mathematics, Physics and Informatics, University of Gdańsk, Wita Stwosza 57, 80-308 Gdańsk, Poland

Correspondence: pawel.rochowski@ug.edu.pl

**Abstract**

The idea behind the research presented is based upon apparently contradictory experimental results obtained here by means of photoacoustics modalities for the same drug donor/acceptor membrane system, serving as a surrogate for a transdermal delivery system. The first modality allowed for the monitoring of the total amount of mass uptake ($m(t)$-type data), while the second technique allowed for the quantification of time-dependent concentration distribution within the acceptor membrane ($c(x,t)$-type data). Despite of a very good agreement between the $m(t)$ data and the 1$^{st}$-order uptake fitting model (standard Fickian diffusion with constant source boundary condition), the standard approach failed during the $c(x,t)$ data analysis. The results led to the analysis of the interfacial transfer contribution to the overall mass transfer efficiency, which eventually allowed to question reliability of the $m(t)$ data analysis for the determination and quantification of the mass transport parameters. A more detailed analysis of the $c(x,t)$ by means of the newly introduced *transport rate number* parameter revealed, that the mass uptake by the acceptor is almost equally influenced by interfacial and bulk transport processes. The analyses performed were translated into a model-free characteristic times, i.e. parameters common for any of the model scheme used.

1. Introduction

Studying the mechanisms of drug membrane transport provides valuable information on the membrane (skin) barrier functions, as well as points at new strategies for efficient drug delivery, which is relevant for medical and cosmetic applications. Since the pioneering work on the membrane transport by Daynes, and followed by studies on the cutaneous membrane delivery published around 60 years ago, there is a strong agreement that the drug passive skin penetration is due to the Fickian diffusion process [1–3]. This fundamental idea of the transdermal transfer origin was confirmed on many occasions simply by stunning similarities between the experimental drug absorption/release curves and the Fickian-type models predictions (or its simplifications, like Higuchi or first-order release models), making the diffusion-based release models a cornerstone of modern pharmacological analysis.

The last decades brought progress in the development of new mathematical approaches and tools for the transdermal drug delivery quantification and modelling; a simple transport models were extended with drug sorption/metabolism/decay kinetic terms, porosity/composition effects, various geometrical conditions, transfer through multi-layered systems, temperature effects etc., but almost always point at Fickian diffusion as a primary (or rate-limiting) transport process [4–9]. It is clear that every mathematical model of a physical phenomenon needs justification and experimental validation. In case of transdermal delivery studies, mathematical models are usually validated upon, so called, drug absorption curves (profiles), which provide information on the total amount of mass (drug) absorbed vs time. Appropriate experiments involve f.e. studies on the donor-to-acceptor transfer kinetics of tracer molecules (dyes or radionuclides, depending on the detection scheme) in the Franz diffusion cell systems [10]. A far less popular experimental techniques involve depth profiling modalities, which allow to scan the penetrant amount in the subsequent strata of a sample during the penetration process [11]. The difference between results offered by the methods is crucial; while the depth profiling techniques provide information on the time dependent penetrant concentration distribution within a system (membrane) - the $c(x,t)$-type data for 1D system, the diffusion cell-based methods yield the $m(t)$-type data, which, in some cases, can be understood as the penetrant concentration integral over a space domain of interest (f.e. $m(t) = \int c(x,t)dx$). Due to the $c(x,t)$-type data processing, at least some of information on the transport mechanisms originally encoded in the time-dependent penetrant distributions can be averaged out, and so may remain undetermined by means of the $m(t)$-type data analysis. The issue also includes the mass release studies (the amount of mass released by an object vs time), as the kinetics constants of drug release (from donor) and drug absorption (by acceptor) must be equal.

The approaches for the absorption/release data analysis usually involve dedicated empirical and semi-empirical formulas/mechanistic models for the determination of intrinsic system characteristics (like diffusion coefficient) under various conditions (f.e. temperature). The latter ones are often given in PDE form, tailored for the sorption data quantification (i.e. PDE to ODE conversion) [12–16]. Both the semi-empirical formulas and the *tailored* mechanistic models are based upon certain approximations of underlying physics and chemistry. If the system characteristics are not properly recognized prior to detailed numerical processing, adoption of *any* model may lead to the data analysis failure [17]. On the other hand, even if some presumptions on the system behaviour can be done, a more detailed analysis on the system properties may be limited by experimental data character (i.e. the averaging problem present in the $m(t)$-type data).

The paper is focused on the mass transfer characterization (in the theoretical and experimental sense) of an ordinary drug (donor)/membrane (acceptor) system, serving as a surrogate for transdermal delivery system. In specific, the role and impact of bulk and interfacial processes on the overall mass absorption kinetics are investigated. Anticipating the results, the theoretical approaches for the mass transport modelling presented were based upon bulk diffusion-based mass transport models studied (mainly) under two types of boundary conditions (BC), *the instantaneous source* (regarded here as the standard approach, where the overall mass transfer is limited by the diffusion process) and *the slowly-equilibrating* BC, as suggested in [18]. The latter one appears to serve as a reasonable approach for the quantification of interfacial processes due to the observable timescale of interfacial relaxations between viscous materials [18–20]. The validation (experimental) data were acquired by means of photoacoustics techniques, namely the photoacoustic spectroscopy (PAS) and the photoacoustic depth-profiling (PADP). Here, for the first time, the drug transfer (from a donor into an acceptor membrane) process is studied by means of $m(t)$-type (characteristic for the Franz diffusion cells experiments) and $c(x,t)$-type data acquired for the same system. The results obtained question one of the paradigm of dermal delivery, namely purely diffusional nature of the passive (i.e. in the absence of additional external stimuli) dermal absorption process: it appears that the drug transfer may be limited by processes occurring at the donor/acceptor interface. It is also demonstrated that the $m(t)$-type datasets analysis 1) may lead to incorrect and misleading quantification of the transport parameters and 2) may be insufficient to study the role of the interfacial processes (affecting the boundary condition character), as the absorption curves exhibit equivocal nature. As such, the need for further depth-profiling studies on viscous delivery systems is highlighted. Eventually, an alternative analysis of the collected $c(x,t)$-type data, based under electric circuit analogue (EC) approach, is presented. The approach relies on the determination of, so called, characteristic time of a system. As such, the approach can be viewed, at least to some extent, as a model free approach. Most of the data analysis presented thorough the paper refers to characteristic times of the absorption processes, which provide an alternative and consistent way for the description of the mass transfer kinetics.

## 2. Theoretical background

**2.1 Basic approaches to the diffusion-guided transport quantification**

Since the pioneering works of of Higuchi and Schleuplein there is a strong agreement that the passive percutaneous drug absorption is mainly due to the Fickian diffusion process. The Fick's laws of diffusion can be given as:

$$\vec{j} = -D\vec{\nabla}c, \tag{1a}$$

$$\frac{\partial c}{\partial t} = -\vec{\nabla}\cdot\vec{j} = D\nabla^2 c, \tag{1b}$$

where Eq. 1a provides the mass flux ($\vec{j}$) as propotional to the product of the diffusion coefficient ($D$) and the permeant concentration gradient ($\vec{\nabla}c$), while Eq. 1b describes the permeant accumulation/depletion rate in a certain volume as proportional to the local curvature of concentration gradient. It can be pointed out, that the righ-hand side of Eq. 1b holds for isotropic systems of constant diffusivity.

A mathematical description of the permeant concentration evolution requires characterization of the initial and boundary conditions related to the surroundings properties and interfacial behaviour. The latter ones are picked to match the experimental conditions, but are often simplified in terms of underlying physics which, in many cases, allows to describe permeation process by a closed-form and elegant mathematical expressions. These expressions can be further abridged if the permeation behavior analysis is limited only to a certain time regime. An example can be given by a pharmacologically relevant 1D mass transport problem described in Crank's textbook (Section 4.3.2 in [21]), i.e. diffusion from a time-independent source into a plane sheet of thickness $2l$. Assuming the sheet ($-l < x < l$) is initially free of the permeant, solution of Eq. 1b under $c(-l,t) = c(l,t) = c_0$ yields expression in the trigonometrical series form:

$$\frac{c}{c_0} = 1 - \frac{4}{\pi}\sum_{n=0}^{\infty}\frac{(-1)^n}{2n+1}\exp\left[\frac{-D(2n+1)^2\pi^2 t}{4l^2}\right]\cos\left[\frac{(2n+1)\pi x}{2l}\right], \tag{2a}$$

which for small times can be approximated by:

$$\frac{c}{c_0} = \sum_{n=0}^{\infty}(-1)^n erfc(\frac{(2n+1)l-x}{2\sqrt{Dt}}) + \sum_{n=0}^{\infty}(-1)^n erfc(\frac{(2n+1)l+x}{2\sqrt{Dt}}). \tag{2b}$$

Eqs 2 allow for the theoretical predictions on the permeant time-dependent concentration distributions, and can be further employed for the qunatification and analysis of experimental data obtained by means of profilometric techniques. In many practical cases, however, a vital information on the drug uptake is related to the total amount of the drug adsorbed by a system (skin, membrane etc.) rather than the drug spatiotemporal concentration profiles. As such, by assuming the total amount of substance inside a membrane ($-l < x < l$) at any time is $m(t) = \int_{-l}^{l} c \, dx$, the evolutions of $m$ (with concentrations as in Eqs 2) are given as [21]:

$$\frac{m(t)}{m_{eq}} = 1 - \sum_{n=0}^{\infty} \frac{8}{(2n+1)^2 \pi^2} \exp\left[\frac{-D(2n+1)^2 \pi^2 t}{4l^2}\right], \quad (3a)$$

$$\frac{m(t)}{m_{eq}} = 2\left(\frac{Dt}{l^2}\right)^{\frac{1}{2}} \left(\pi^{-\frac{1}{2}} + 2\sum_{n=1}^{\infty}(-1)^n ierfc\left[\frac{nl}{\sqrt{Dt}}\right]\right), \quad (3b)$$

where $m_{eq}$ denotes the asymptotic value of $m(t)$ for large times. Eqs 3 provide a basis for two well-known mass absorption models; in particular, by limiting the summation over *n* to the first term only (*n=0*), Eq. 3a yields the first-order absorption model (the mass absorption process is expressed by a simple monoexponential growth function):

$$\frac{m(t)}{m_{eq}} = 1 - \frac{8}{\pi^2} \exp\left[\frac{-D\pi^2 t}{4l^2}\right] = 1 - F \exp[-\gamma t]. \quad (4a)$$

By droping the summation in Eq. 3b one obtains the Higuchi equation:

$$\frac{m(t)}{m_{eq}} = 2\left(\frac{Dt}{\pi l^2}\right)^{\frac{1}{2}} = kt^{\frac{1}{2}}. \quad (4b)$$

Eqs 4 provide basic mathematical models for the quantification of drug absorption into a single homogeneous system. In many practical cases, however, one deals with multilayer systems (i.e. in the Franz diffusion cells systems, where the permeate migrate from donor → membrane → acceptor, or dermal patches, where drugs migrate from donor → skin layers → blood vessels), or porous/composite systems (i.e. multiphase structure of skin – *the brick and mortar model*, and, similarly, skin-mimicking semipermeable membranes). Lately, a new approach for the modelling of mass transport through composite systems, based on analogies between the mass and charge transfer (Fick's and Ohm's laws), has been proposed [22]. In general, subsequent compartments of the mass transport system are characterized simply by their volumes and resistances (inversely proportional to conductivities). Consequently, the mass transport can be modelled in analogy to electric circuits consisting of capacitors and resistors pairs, i.e. by means of Kirchhoff's voltage (the concentration analogue) or current (related to the mass flux) laws. For a system consisting of a time-independent donor and an acceptor, the evolution of the penetrant amount in the acceptor phase is given by:

$$\frac{m(t)}{m_{\text{eq}}} = 1 - \exp[-t/RC] = 1 - \exp[-t/\tau], \tag{5}$$

where $R$ and $C$ stands for the acceptor resistance and capacitance (which is a function of the acceptor volume $V$; for simplicity we take $C \equiv V$), respectively, while $\tau$ is the exponential-process characteristic time (i.e. time required for the system to reach $m(\tau)/m_{\text{eq}} \approx 63\%$ of its asymptotic value). It is clear that Eq. 5 has similar mathematical form to Eq. 4a. In practice, however, an additional geometrical-correction factor has to be included to meet required convergence to the Fick's model predictions (i.e. Eq. 3a). In case of planar geometry, Eq. 5 can be rearranged to:

$$\frac{m(t)}{m_{\text{eq}}} = 1 - 0.86 \exp[-t/\tau^{EC}] = 1 - 0.86 \exp[-2.62Dt/l^2]. \tag{6}$$

where $\tau^{EC} = l^2/2.62D$ stands for the electric circuit-based mass transfer model characteristic time for plane-sheet geometries. There are, at least, two advantages of the electric-circuit (EC) based approach; first, the permeation process is characterized by the characteristic time $\tau^{EC}$ and optionally by the diffusion coefficient (for a specific case of purely diffusional transport process a direct relation between the parameters is given by $\tau^{EC} \propto l^2/D = \tau_D^n$; note, that $\tau_D^n$ is a more natural definition of the characteristic diffusion time, as it is based on the system characteristic parameters only). The characteristic time can be understood as a certain measure of the hindrance that molecules encounter during the permeation, without referring to any particular transport mechanism. As such, the approach can be viewed as a model-free approach, which seems to be a reasonable choice when the transport mechanisms cannot be directly identified. Second, the quantitative analysis of composite systems (porous, multicompartmental, multilayered etc.) is based on the well-known principles (Kirchhoff's laws), the calculus is rather straightforward, and does not require to solve any sophisticated and cumbersome mathematical problems or adaptation of any transport model. Another possible advantage includes generalization of the simple exponential form (Eq. 5) to the Weibull form, which in case of the mass release (note that EC approach predicts a time symmetry between the mass absorption and release, or, in other words, the characteristic time for absorption and release is the same) is given by [23]:

$$\frac{m(t)}{m_{\text{eq}}} = \exp[-(t/\tau)^b]. \tag{7}$$

The Weibull form allows for the analysis of stretched experimental data (in case of purely diffusion models, this would indicate a certain dispersion of the diffusion coefficient, f.e. due to the diffusion coefficient concentration dependence or variations of the systems characteristic lengths). It can be noticed, that the stretched exponential functions are commonly used in the of relaxation data analysis [24].

## 2.2 Photoacoustics

The photoacoustics modalities are based upon the detection of gas pressure variations following the absorption of modulated light (or more general, energy flux) by a sample. A standard approach for the quantification of the PA signal is based upon the Rosencwaig-Gersho's thermal piston model [25]. In brief, the signal generation consists of subsequent steps: (I) perturbation of the temperature field in the sample due to the Lambert-Beer-type incident light absorption; (II) evolution of the perturbed temperature field governed by a set of Fourier-Kirchhoff equations for the sample, surrounding gas and backing material; (III) determination of gas pressure variations (due to periodic and adiabatic heat transfer from the sample to the surrounding gas) in the vicinity of the sample ($\delta P$), recorded by a detector (the PA signal $S \propto \delta P$) [11].

A photoacoustic response of a sample depends on an interplay between the sample physical parameters, i.e. its thickness ($l$), the thermal diffusion length ($\mu_\alpha = 1/\alpha$, where $\alpha$ stands for the sample thermal diffusion coefficient) and the optical absorption length ($\mu_\beta = 1/\beta$, where $\beta$ stands for the Lambert-Beer optical absorption coefficient). When measured against constant modulation frequency and varying wavelength $\lambda$ (the PA spectroscopy mode), in the absence of saturation effects ($\mu_\alpha < \mu_\beta$) the photoacoustic response can be given by:

$$\delta P(\lambda) = A\, I(\lambda)\eta(\lambda)\beta(\lambda), \tag{8}$$

where A is the apparatus constant, $I(\lambda)$ is the incident light intensity, and $\eta(\lambda)$ stands for the non-radiative conversion efficiency. Due to the $\delta P(\lambda) \propto \beta(\lambda)$ relationship, the PA spectra exhibit similarities to the standard absorption spectroscopy data. A great advantage of the PA method over the standard absorption spectroscopy routines involves no requirement on the sample preparation. As such, the pigment behaviour inside a membrane (permeation, chemical reactions etc.) can be monitored in-vivo, without any need for an additional mechanical treatment.

The second photoacoustic modality considered involves a situation, where the PA response is measured against varying modulation frequency and pseudo-monochromatic wavelength conditions (the PA depth-profiling mode). Then, the relation between the photoacoustic response and the modulation frequency can be given by:

$$\delta P(\lambda) \propto A\, \eta(\lambda)/f^n, \qquad (9)$$

where $n \in\, <1, 1.5>$, depending on the thermal and optical properties ($\mu_\alpha$, $\mu_\beta$) of the sample [26]. A crucial property of the PA techniques mentioned is a relation between the sampling depth $L$ (a specific sample thickness contributing to the PA signal, taken to be equal to the $\mu_\alpha$) and the incident light modulation frequency:

$$L = (\alpha/\pi f)^{0.5}. \qquad (10)$$

By benefiting from Eq. 10, it is possible to acquire PA signal corresponding to subsequent layers of a material simply by varying the light modulation frequency. Furthermore, considering the PA signal magnitude proportional to the pigment concentration via $\beta(\lambda) = \epsilon(\lambda)c$ relationship ($\epsilon(\lambda)$ stands for the molar absorptivity), the PA signal magnitude is also a measure of the pigment amount from the surface to the sampling depth [27]:

$$m(t) = \int_0^L c(x)dx. \qquad (11)$$

As such, by performing multiple PA scans at a fixed frequency and over a certain period of time it is possible to record the mass absorption/release curves (data in the $m(t)$ form). If the experiments are performed at various frequencies (various sampling depths), it is possible to evaluate local pigment concentration ($c(x,t)$-type data) by subtracting signal from two neighboring layers and dividing by the layer thickness [11].

### 3. Materials and methods

The experiments performed involved photoacoustic detection of dithranol (in pharmaceutical form) absorption by dodecanol-collodion (DDC) membranes. The system has been previously identified as a reliable candidate for the photoacoustic proof-of-concept studies due to dithranol high thermal deactivation efficiency $\eta(\lambda)$ and non-overlapping absorption peaks of the drug (~ 360nm), its photoproducts (~ 430 nm), and the membrane (< 280nm). 2% dithranol/Vaseline suspension was delivered by a local pharmacy. Pure dithranol (PN: D2773) was purchased from Sigma-Aldrich. The tissue-mimicking DDC membranes (of thickness of $l = 23 \pm 2$ μm) were synthesized from collodion solution mixed with 1-dodecanol (with final dry mass ratios 2:1), as described earlier [11]. Photoacoustic characterizations of the model drug suspension (in comparison to pure dithranol sample) and the membrane is shown in Fig. 1a.

All the experimental studies were performed using the same apparatus. In particular, the set-up consisted of a 900 W xenon light source (Newport 66921), grating monochromator with adjacent mechanical chopper (Newport 74125), PAC300 photoacoustic chamber (MTEC Photoacoustics) and SR7265 lock-in amplifier. A block diagram of the rig is shown in Fig. 1b. The PA spectra were recorded in the range of 250-500 nm in 5 nm steps and at modulation frequency of 37 Hz. For the depth-profiling studies, nine frequencies in the frequency range of 37-170 Hz were selected. Assuming the thermal diffusivity of the membranes synthesized was $\alpha = 5.57 \cdot 10^{-8}\ m^2\ s^{-1}$ [11], the limiting sampling depths were ~10 – 22 µm from the surface (or ~1 – 13 from the drug/membrane interface).

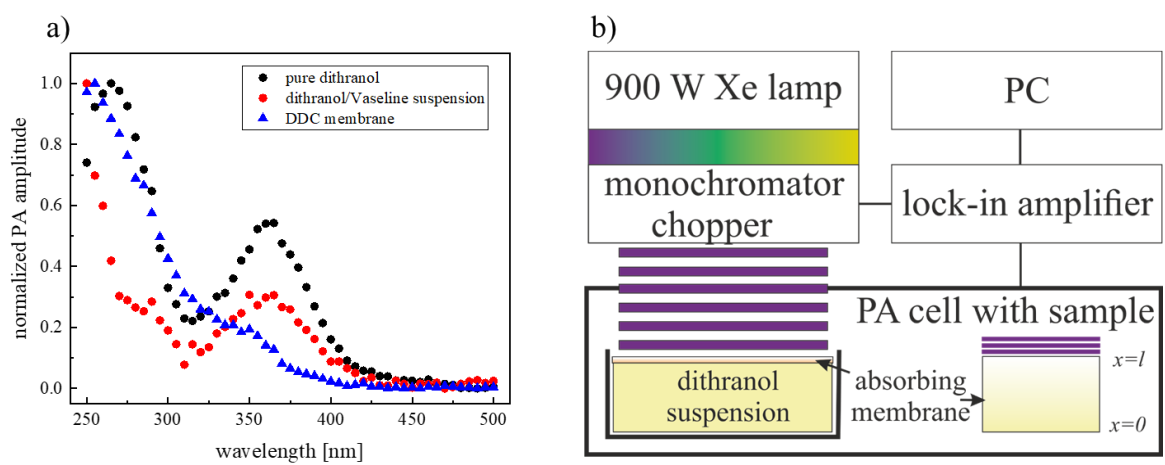

Fig. 1. a) photoacoustic spectra of pure dithranol, the model drug donor phase (dithranol/Vaseline suspension) and the acceptor - DDC membrane; b) – a block diagram of the PA apparatus along with the PA cell loaded with the sample.

The depth-profiling studies for a similar system were performed previously [11], using 1 W 370 nm LED as an incident light source. It may be noted, that a LED source driven by a function generator allows for faster frequency swapping and expands the modulation frequency range. Here, however, all the measurements were performed within a single apparatus to avoid any experimental inconsistencies (i.e. different bandwidths/powers of the light sources, average times for the modulation frequency stabilization).

All the experiments were performed at temperature T=25 ºC.

The numerical analysis were performed using OriginPro 2018 (OriginLab) and Mathematica 12.2 (Wolfram).

## 4. Results and discussion

### 4.1 On the transport rate limiting processes

Let us consider a case of drug sorption by a membrane of characteristic length $l$. The membrane ($x \in (0, l)$), initially free of drug, is assumed to be in a one-side (bottom side as in Fig.1b) full contact with the drug reservoir ($x \leq 0$), i.e. the cross sections of the membrane and the reservoir are equal (1D transport approximation). Also it is assumed, that the average drug concentration in the reservoir remains constant over the membrane drug sorption time period, and is equal $c_0$ (this assumption is valid as long as the capacity of the reservoir is much greater than the capacity of the membrane; as the cross sections of the both compartments are assumed to be equal, the capacities are related directly to the characteristic lengths of the compartments). As such, the diffusion-driven transport problem for $x \in (0, l)$ (the membrane domain) under the membrane constant diffusivity condition takes the form:

$$\frac{\partial c}{\partial t} = -\frac{\partial}{\partial x}\left(-D\frac{\partial}{\partial x}c\right) = D\frac{\partial^2}{\partial x^2}c, \tag{12a}$$

$$c(x, 0) = 0 \text{ for } x > 0, \tag{12b}$$

$$\partial c(l, t)/\partial x = 0, \tag{12c}$$

where the last condition implies no mass flux at the upper membrane interface. Solution of the transport problem (Eq. 12a) requires defining the lower boundary condition, related to the mass transfer from the reservoir (drug donor) to the membrane (acceptor). The solutions to the problem considered by Crank (Section 2.1) were obtained by setting constant concentration at the mass donor/acceptor interface, which physically means instantaneous drug mixing within the donor phase and transfer to the membrane outermost layer adjacned to the reservoir (or, in other words, the membrane transport as a whole is limited by the slowest process, which in the case considered is the membrane diffusion). Similarly, the first type of lower BC for the Eq. 12a can be given by the Dirichlet-type instantaneous source condition: $c(0^+, t) = c_0$ (condition I, $C_I$). The condition is valid in case of high interfacial mass transfers with respect to the membrane bulk diffusion, which leads to instantaneous equilibration at the donor/acceptor interface (the concentration at the boundary of the membrane domain resembles drug concentration in the reservoir). In case of low interfacial transfers the $C_I$ condition must be replaced by the Neumann-type condition reflecting the magnitude of interfacial flux, given by: $j|_{x=0} = \phi(c_0 - c(0^+, t))$ (condition $C_{II}$), where $\phi$ stands for the interfacial mass transfer coefficient. To study the impact of the magnitude of interfacial mass transfer on the membrane sorption kinetics let us introduce a dimensionless transport parameter provided by a ratio of the interfacial mass transfer and characteristic length to the bulk diffusion coefficient, and hereafter referred to as the transport rate number: $N_R = \phi l/D$. The parameter introduced reminds of other dimensionless mass transfer parameter, i.e. the Peclet number ($Pe = vl/D$), which measures an impact of the bulk diffusion and advection (characterized by advection velocity $v$) processes on the mass transport rate, or the Sherwood number, given by the ratio of advective to diffusive mass transfers across a boundary [28]. Alternatively, the transport rate number can be expressed by means of the interfacial mass transfer and diffusion characteristic times: $\tau_\phi = l/\phi$ and $\tau_D^n = l^2/D$, respectively (the "n" superscript indicates, that the characteristic time is defined by the system characteristics only; note, that: $\tau_D^n = 2.62\, \tau^{EC}$). As such, $N_R = \tau_D^n/\tau_\phi$. It can be noted, that for $N_R \gg 1$ the membrane sorption kinetics is limited by the membrane intrinsic property – its diffusivity, while for $N_R \ll 1$ the sorption is limited by the interfacial mass transfer.

It should be emphasized, that even in the presence of high interfacial fluxes the $C_I$ condition may not be valid for *small* times. This is due to the presence of interfacial relaxation process characteristic for a newly created system (here: by adjoining drug donor and acceptor subsystems), which leads to new equilibrium value of the surface excess concentration (the surface excess concentration of a component at the equilibrium is related to the observable difference between the surface and the bulk concentrations of the component) [18]. As such, the $c(0^+, t) = c_0$ ($C_I$) condition (valid for high interfacial fluxes, as explained earlier) should be replaced with a kinetic condition: $c(0^+, t) = \Gamma(t)\delta x_0$, where $\Gamma(t)$ represents the evolution of surface excess and $\delta x_0$ strands for the thickness of the interface layer, while for the $C_{II}$ one obtains: $j|_{x=0} = \phi(\Gamma(t)\delta x_0 - c(0^+, t))$ [18]. Recent studies on the dynamic interfacial tension behaviour between two liquid phases indicate an quasi-exponential character of $\Gamma(t)$ [29]. To keep the overall picture of the membrane sorption clear, we define the third boundary condition ($C_{III}$) in the fast interfacial transfer limit as $c(0^+, t) = 1 - \exp[-\varepsilon t]$, with ε as the interfacial equilibration constant. By definition, $\varepsilon = 1/\tau_{eq}$, where $\tau_{eq}$ is the interfacial equilibration time constant. Eventually, the transport rate number becomes: $N_R = \tau_D^n/(\tau_\phi + \tau_{eq})$, and $N_R = \tau_D^n/\tau_{eq}$ in the fast interfacial transfer limit approximation.

Exemplary sorption curves for the diffusion-driven transport models (Eqs 12) under $C_{II}$ and $C_{III}$, additionally compared to the classical $C_I$ case, are shown in Fig. 2. The simulations were performed for: material diffusivity $D = 1$ [a.u.], material thickness $l = 1$ [a.u.], transport rate number $N_R$ in the range of 0.01 – 100. Additionally, the time (x-axis) values for each process was divided by the characteristic time of the process (understood here as a specific time when $m(t)/m_{eq}$ reaches 63% of its equilibrium value, as in case of an exponential process; for the $N_R$'s considered here, the characteristic times were ($N_R/\tau$): 0.01/111, 0.1/10.1, 1/1.39, 10/0.43, 100/0.39), to accent similarities between the profiles. From Fig. 2a ($C_I$ vs $C_{II}$) it can be noticed, that regardless of the $N_R$ magnitude, the absorption curves are almost undistinguishable. In other words, for the $C_I$- vs $C_{II}$-type curves considered it is cumbersome to point at a process dominating the mass absorption (bulk absorption vs interfacial transfer) by the curve shape (or curvature) analysis only. On the contrary, the evolution of concentration profiles ($c(x,t)$ data) of $C_{II}$ for relatively low $N_R$s (see Fig. S1a in the Supplementary materials section, where 1 (upper row) is for $N_R = 10$, while 2 (middle row) - $N_R = 1$, Fig.3c – solution of Eqs 12 under $C_I$) exhibit evidently different characteristics than the $C_I$, both close to the interface (quasi-exponential growth vs constant concentration) and in the bulk (for $C_{II}$: subtle concentration gradients along sample thickness for consecutive times). Fig. 2b demonstrates absorption curves for $C_I$ vs $C_{III}$ case (corresponding $c(x,t)$ for $N_R = 10$ and $N_R = 1$ are shown in Fig. S1b). The curvature differences between the $C_I$ and $C_{III}$ profiles are more noticeable compared to the previous case, especially for small and medium $N_R$'s (in fact, this is expected feature, as for large surface equilibration constants the surface concentration seen from the experimental time window perspective mimics the constant concentration condition: for $c(0^+, t)$, $\lim_{\epsilon \to \infty}(1 - \exp[-\varepsilon t]) = 1$). It can be noticed, that for decreasing $N_R$'s the $m(t)/m_{eq}$ profiles tend to exponential character.

It is worth to consider whether $C_{III}$-type processes can be distinguished from $C_I$-type by analysis of $m(t)/m_{eq}$ experimental data. To explore the issue 24.000 artificial absorption profiles following Eqs 12 and $C_I$ (for $D = 1$ [a.u.], $l = 1$ [a.u.]) condition were generated. In general, the data base consisted of subsets, characterized by distinct number of experimental points (3, 5, 10 or 20 points) *measured* with a certain *precision* upper limit (1%, 3% or 5%), and randomly distributed within a certain time interval (short times: $t\epsilon(0; \tau)$ with $m(\tau)/m_{eq} = 0.63$, and long times: $t = t\epsilon(0; 5\tau)$ with $m(5\tau)/m_{eq} \approx 0.99$). As such, a single set (specific number of points, precision, time interval) consisted of 1000 various profiles. Exemplary profile (10, 5%, $5\tau$) along reference absorption curve (numerical solution for Eqs 12 under $C_I$ BC) is shown in

Fig. S2. All the data sets were fitted by means of three models, provided by Eqs 4 (exact solution, Higuchi) and 6 (monoexponential EC model), to provide apparent values for the diffusion coefficient (results are collected in Tab. S1). The results of the analysis indicate, that even for the highest precision (1%) and considerably narrow experimental window - $t\epsilon(0;\tau)$, it would be hard to differentiate between the $C_I$ and $C_{III}$ conditions, as the exponential and exact model based analyses of the data sets yield similar results in terms of the free parameter value (within the uncertainty range). As exponential functions (f.e. Eqs 4a, 5 and 6) are considered as an approximation for the exact solution (i.e. Eq. 3a) valid for long times one can point at very narrow experimental window ($t \ll \tau$) to discriminate between origins of the interfacial processes. This, however, requires rapid experimental methods and protocols. An alternative way is provided by methods allowing for the permeant detection in the system bulk.

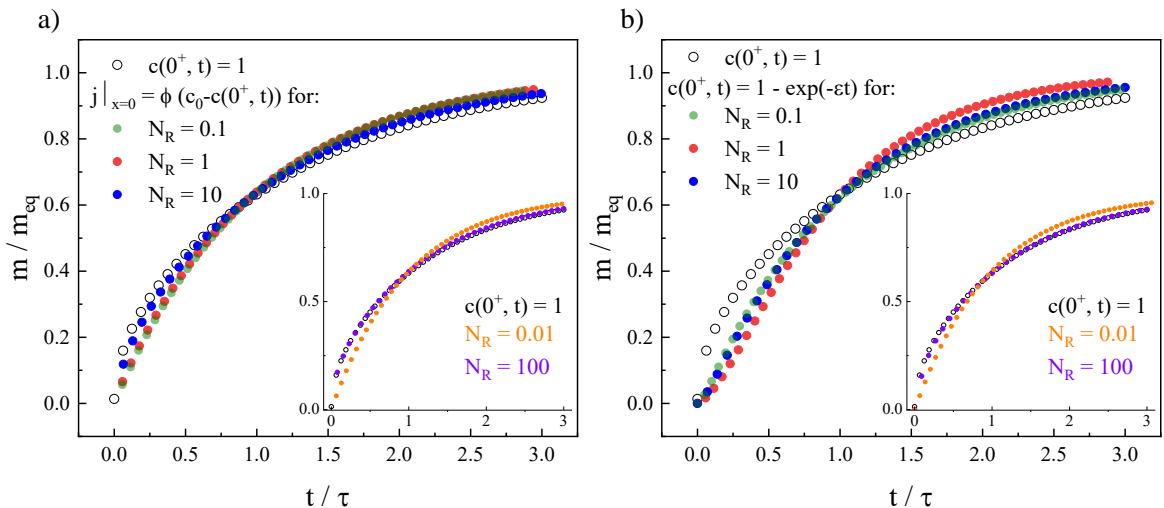

Fig. 2 Sorption profiles for a finite membrane of diffusivity $D = 1$ and thickness $l = 1$ in a one-side contact with a drug reservoir. a) donor/acceptor boundary condition given by interfacial flux of magnitude $N_R$ ($C_{II}$ condition), b) the boundary condition described by exponential relaxation function with the equilibration constant of magnitude $N_R$ ($C_{III}$ condition).

### 4.2 PAS experiment and the lumped data set

This and the following sections are dedicated to showcase experimental evidences for the equivocal nature of the mass absorption curves.

The first membrane transport experiment performed involved spectroscopic detection of the pigment (dithranol) content evolution, $m(t)$, in the DDC membrane. The modulation frequency of 37 Hz (L = 22 μm) was chosen to sample the membrane maximal thickness possible without overstepping the drug/membrane interface boundary. The whole experiment took around 2 hours, with a single PA spectra (250-500 nm) recorded during a time span of 14 minutes. The results of the experiment are shown in Fig. 3, where (a) presents the normalized PA response (with respect to carbon black) as a function of wavelength and time (data as recorded during the experiment), and (b) - the same data but in the spectra series representation (to enhance the data clarity). Two observations can be immediately made. First, the dithranol-characteristic 360 nm band (see Fig. 1a) increases with time, however, the increment rate slows down for the longer times. Second observation refers to the stability of the drug. It is known that dithranol is relatively unstable and undergoes autooxidation to quinone derivatives and dimers (and, eventually, further polymerization), with danthrone and dithranol dimer as the main oxidation products in aprotic and protic solvents, respectively. The dithranol degradation rate is typically enhanced in the presence of light, increased temperature and exposure to molecular oxygen [30].

During our previous dithranol photostability and transport studies by means of photoacoustics techniques (with strong LEDs as light sources) it was shown, that the drug remained photostable in dodecanol solution, however, underwent photodegradation to two products during the membrane penetration process. The degradation products were characterized by PA bands at ~400 and ~430 nm (in fact, the 430 nm band is the spectroscopic fingerprint of danthrone molecules), which clearly deformed the dithranol PA band [11]. Here, by comparing the final spectra in Fig. 3b with the model dithranol spectrum in Fig. 1a it can be concluded, that no additional strong bands in the 380 – 500 nm region appeared (in comparison to Fig. 2(d-e) in [11]), which suggest significant chemical stability of the drug, at least in the time window considered and under relatively weak Xe lamp light source irradiation. Eventually, the governing transport equation considered here (Eq. 12a) needs no further extensions for depletion terms (reaction terms) and preserves its simplest form.

The PA signal evolution from Fig. 3b at a fixed wavelength of $\lambda = 360$ nm is shown in Fig. 3c (red points). In the absence of the drug degradation processes the time-dependent PA signal behaviour is expected to obey the diffusion-based model. Let us assume the classical approach to the problem involving the $c_I$-type BC, with $m(t)/m_{eq}$ represented by one of the Eqs: 3a, 4a or 5-6. By applying the best-fit procedure of the monoexponential model (Eq. 6) to the experimental data one obtains characteristic time of the penetration process $\tau_{PAS}^{EC} = 0.42 \pm 0.02$ h ($\tau_{PAS}^{n} = 1.10 \pm 0.05$ h), which can be further translated into the diffusion coefficient as the relation holds: $D = l^2/2.62\tau_{PAS}^{EC}$. As such $D_{PAS} = (1.45 \pm 0.21) \cdot 10^{-9}$ cm$^2$ s$^{-1}$. The best-fit curve is presented in Fig. 3c. A very good convergence of the best-fit to the experimental data can be noticed, with the goodness-of-fit parameter $R^2 = 0.99$.

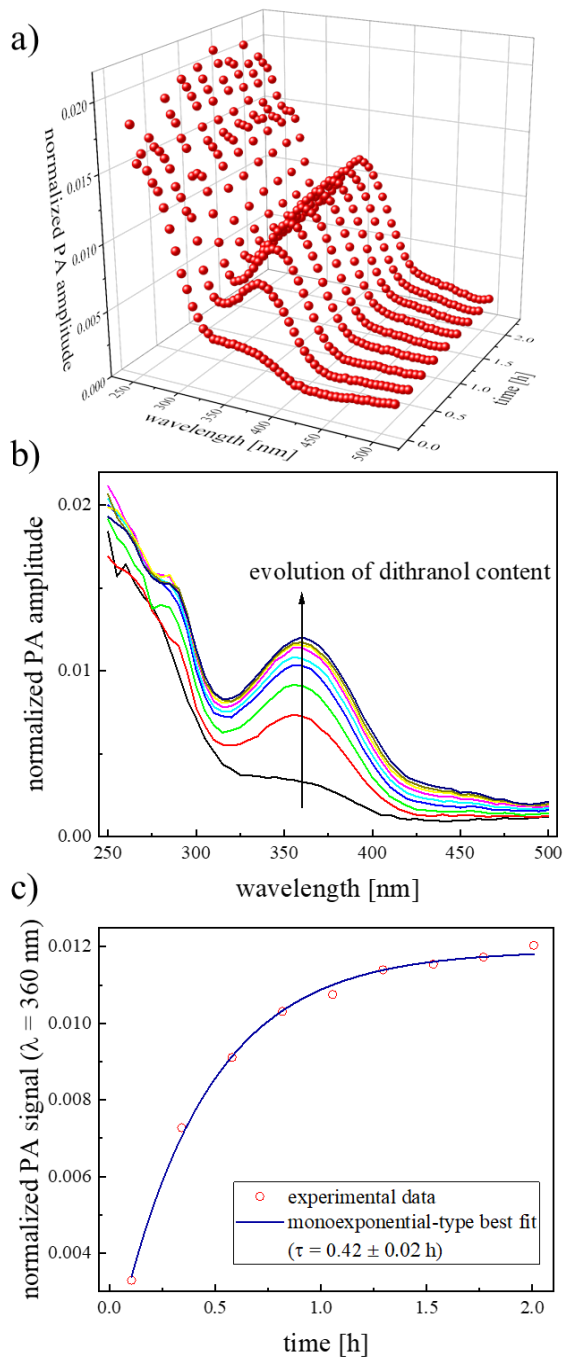

Fig. 3 a) time-resolved PA spectra of DDC membrane in contact with dithranol suspension; b) a simplified data representation for the membrane transport kinetics analysis; c) signal evolution from b) at the fixed wavelength of $\lambda = 360$ nm (red points) and the monoexponential (Eq. 6) best fit curve.

## 4.3 PADP experiment and the distributed data set

The second membrane transport experiment involved PADP studies on a very similar system (i.e. same chemical composition, geometry, PA apparatus etc.) as during the spectroscopic experiment; the only difference involved the character of the incident light – fixed wavelength (360 nm) and varying modulation frequencies. The lowest modulation frequency was picked to match the modulation frequency of the spectroscopic experiment, while the highest one was the maximal obtainable on the apparatus configuration used. Raw data (evolution of the PA signal for subsequent modulation frequencies) collected during the experiment are collected in Fig. S3a. To convert the $S(f,t)$ data into the $c(x,t)$ data (the raw PA signals into the drug concentration evolution picture) the method extensively described in the Section 4.4 of [11] and briefly here (last paragraph in the Theoretical background) was used. The drug concentration profiles, normalized to unity, are shown in Fig. S3b.

The experimental data set from Fig. S3b was analysed by means of Eqs 12 under $C_I$ and $C_{III}$ conditions (the best-fits were obtained by the NonlinearModelFit procedure with Levenberg-Marquardt algorithm). Results of the best-fit procedures applied to the data are shown in Fig. 4. In the first case (the instantaneous source condition, one fitting parameter, the diffusion coefficient, Fig. 4a), the best-fit diffusion coefficient was found to be $D_{C_I} = (1.35 \pm 0.03) \cdot 10^{-9}$ cm$^2$ s$^{-1}$, which is comparable to the value found during the spectroscopic data analysis ($D_{PAS} = (1.45 \pm 0.21) \cdot 10^{-9}$ cm$^2$ s$^{-1}$). The goodness of fit parameters are: $R^2 = 0.98$ and the estimated variance (residual sum of squares divided by the number of degrees of freedom; in general, the lower variance, the better fitting quality) $V_{C_I} = 7.38 \cdot 10^{-3}$. An interesting insights into the fitting quality can be obtained by analysing the best-fit residuals (the difference between the acquired data and best-fit); the best fit residuals are shown in Fig. 4b. The biggest discrepancies can be noticed at the initial stage of the transport process (up to ~30 min, point number < 50), especially for the highest scanning depths (close to the drug donor/membrane interface). After the 30 minutes period the discrepancies magnitudes rarely exceeds 5%. It can be noted, that the experimental data close to the donor/acceptor boundary reflects quasi-exponential behaviour, characteristic for permeation processes affected by interfacial hindrance effects (see Fig. S1).

The second scenario considered (Fig. 4c-4d) involved two fitting parameters (the diffusion coefficient and the interfacial equilibration constant), which were found to be $D_{C_{III}} = (4.84 \pm 0.41) \cdot 10^{-9}$ cm$^2$ s$^{-1}$ and $\varepsilon = (5.88 \pm 0.25) \cdot 10^{-4}$ s$^{-1}$ ($\tau_{eq,C_{III}} = 1692 \pm 72$ s $= 0.47 \pm 0.02$ h). By employing Eq. 6 it is possible to determine characteristic times of bulk diffusion process, which are $\tau_{C_{III}}^{EC} = 0.13 \pm 0.01$ h or $\tau_{C_{III}}^{n} = 0.33 \pm 0.03$ h (in comparison to $\tau_{C_I}^{EC} = 0.45 \pm 0.02$ h and $\tau_{PAS}^{EC} = 0.42 \pm 0.02$ h). The transport rate number for the system considered $N_R = \tau_{C_{III}}^{n}/\tau_{eq,C_{III}} \approx 0.7$, which indicates mixed transfer regime, i.e. both interfacial and bulk processes significantly influence absorption kinetics. The goodness of fit parameters are $R^2 = 0.99$ and the estimated variance $V_{C_{III}} = 1.64 \cdot 10^{-3}$. As $V_{C_{III}} < V_{C_I}$ one can expect better experimental data fitting quality for the $C_{III}$ approach. Indeed, the best-fit residuals for the scenario (Fig. 4d) reflect better fitting quality; the majority of residuals at the initial stage of the process are lower than 6%, and drop below 5% for the following stages.

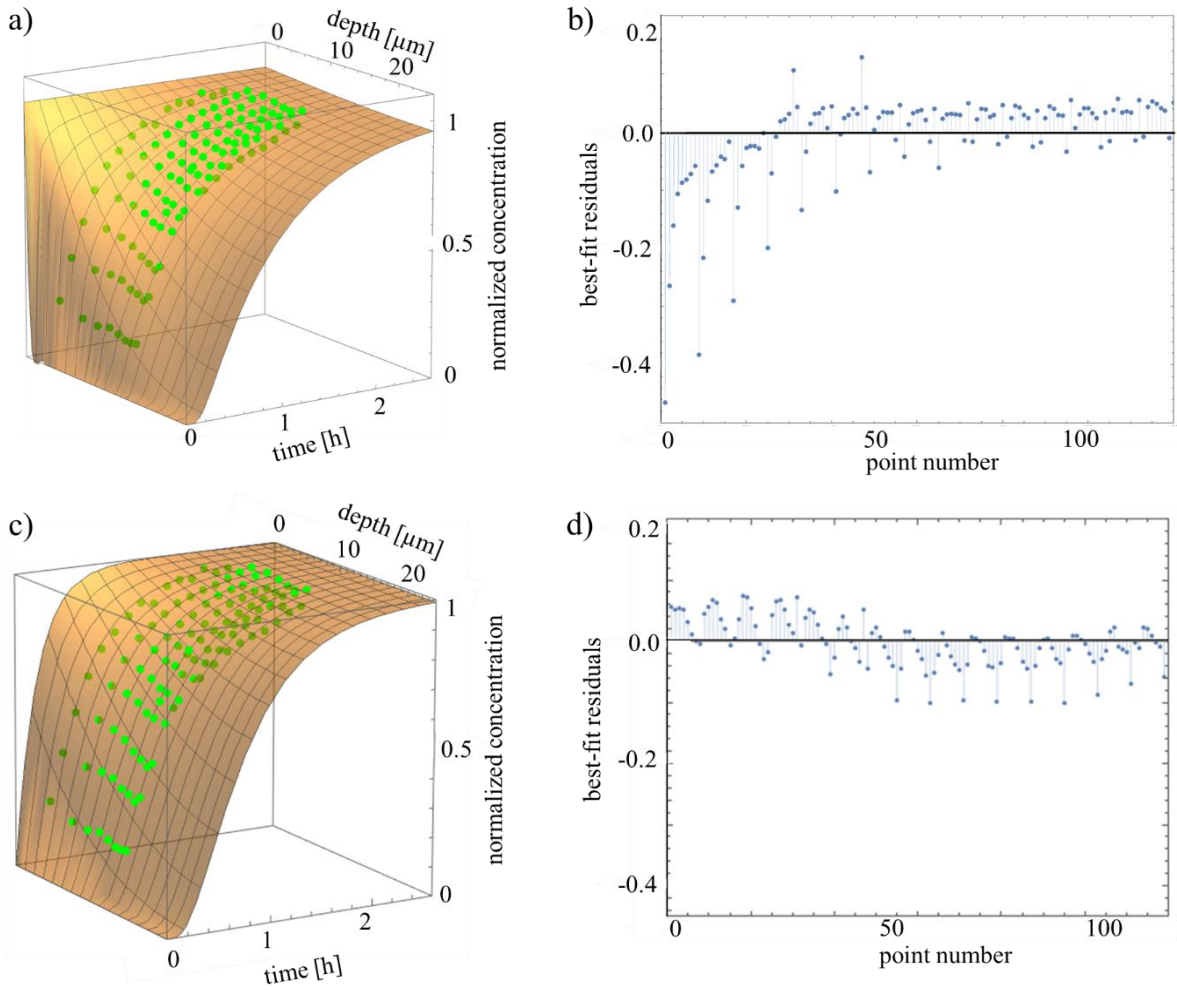

Fig. 4. Results of the best-fit procedure of the diffusion model under $C_I$-type (a-b) and $C_{III}$-type BCs (c-d) applied to the data from Fig. S3b: experimental points vs the best-fitting plane (a and c) and the best-fit residuals (b and d).

It can be noticed, that the both experiments, PAS and PADP, provided experimental data on the PA signal evolution following the $\lambda$ = 360 nm beam absorption at f = 37 Hz, which can be analysed by means of the absorption $m(t)/m_{eq}$-type model. The experimental data (green full circles) from the PADP experiment along with the exponential best-fit (blue line), confronted against the data from the PAS experiment (Fig. 3c, red circles), are shown in Fig. 5. It can be noticed, that the lumped PADP and PAS experimental data follow very similar trend as predicted by the standard approach (i.e. instantaneous source condition). The best-fit characteristic time for the lumped PADP data, $\tau_{PADP}^{EC} = 0.45 \pm 0.01$ h, matches its PAS counterpart ($\tau_{PAS}^{EC} = 0.42 \pm 0.02$ h) within the uncertainty range, and the best-fit result obtained during the distributed dataset analysis under $C_I$ condition: $\tau_{C_I}^{EC} = 0.45 \pm 0.02$ h.

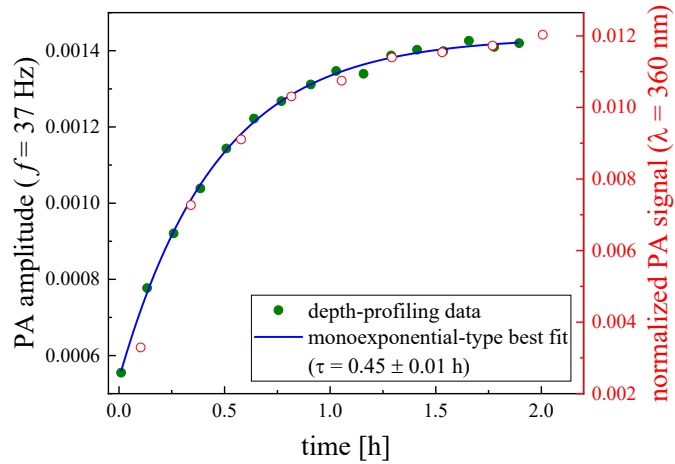

Fig. 5. PADP signal evolution from Fig. S3 at the modulation frequency of 37 Hz (green circles) along with the best-fit curve (blue) given by Eq. 6, and PAS data from Fig. 3 (red circles).

Summarizing, the analysis of the lumped data sets ($m(t)$-type data) from PAS and PADP experiments led to comparable findings on the system transport properties (i.e. $D \approx 1.4 \cdot 10^{-9}$ cm$^2$ s$^{-1}$, which can be optionally translated into characteristic time of the permeation process), with very reasonable and convincing data fitting quality. The results stand in line with the best-fit diffusion parameter obtained during the analysis of a distributed data set ($c(x,t)$-type data) by means of the Fickian model with $C_I$-type (instantaneous source) boundary condition. The qualitative analysis revealed discrepancies between the experimental data and the model predictions especially in the donor/acceptor interface region. The PADP data analysis based upon the Fickian model under $C_{III}$ BC (which refers to the time-dependent interfacial equilibration between donor and acceptor phases) led to a better fitting quality (estimated variance of the fit was 3-times lower compared to the $C_I$-based approach), but also revealed the diffusion coefficient to be ~3.5 times higher compared to the estimate provided by lumped data set analysis. Eventually it was found, that the mass transport process is significantly affected by interfacial and bulk transfer rates ($N_R \approx 0.7$).

**4.4 The magnitude of the classical approach failure**

In the previous sections it has been demonstrated, that the mass uptake rate can be limited by surficial processes, but the interfacial hindrance effects might not be easily quantifiable upon the analysis of mass absorption curves only. As such, absorption curves encode information on the barrier properties of an absorbing system as a whole, disregarding characteristics of particular regions. This loss of region-specific information on the system barrier properties distort overall picture of mass transfer scheme, and may lead to inaccurate predictions for more complex structures, for example involving mass transfer through multi-layered systems.

To estimate the magnitude of *the classical approach failure*, understood as the discrepancy between an apparent diffusion coefficient of the $C_I$-type process and the actual diffusion coefficient in the presence of interfacial hindrance effect modelled by means of the $C_{III}$-type process (considered as a plausible class of the processes, as shown in the photoacoustic experimental data), artificial data sets have been modelled with use of Eqs 12 ($D = 1$ [a.u.], $l = 1$ [a.u.]) and $C_{III}$ boundary conditions (with $N_R$ spanning from 0.01 to 100), and integrated over the *x* domain to provide absorption profiles (each data set involved ~30 points in the range of $t \in (0; 3\tau)$). The absorption profiles were then fitted by means of a standard exponential fitting scheme under $C_I$ BC (Eq. 6) with only $D$ as a fitting parameter (the apparent diffusion coefficient). Results of the best-fit procedure are shown in Fig. 6. The standard errors of the estimates (diffusion coefficients) were less than 2%, and were omitted in the figure. The goodness of fit $R^2$ parameters were always greater than 0.98.

It can be noticed, that, on the one hand, the quantification of the absorption data by means of the classical diffusion model yields a very convincing convergence (high $R^2$). On the other hand, the model generally fails for the actual material constants value estimation; a reasonable results are obtained only for diffusion-controlled problems ($N_R$>>1). Due to high fitting accuracy and the data/model convergence, it appears that it is not possible to differentiate between the diffusion-controlled and interface transfer/relaxation-controlled processes by referring to the absorption profile curvatures only, at least upon data collected for times covering most of the sorption process (here: $t \in (0; 3\tau)$ and so $m(3\tau)/m_{eq} \approx 0.95$).

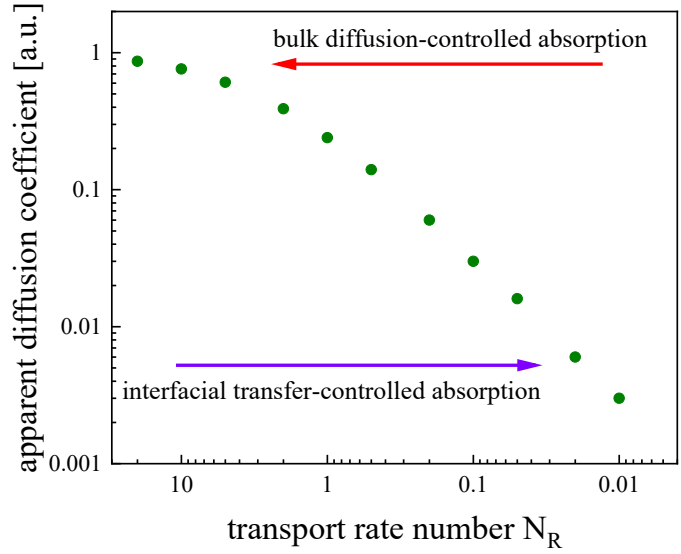

Fig. 6 The apparent (best-fit) diffusion coefficient as a function of the transport rate number ($N_R = \tau_D^n / \tau_{eq}$) for an absorbing system characterized by $D = 1$ [a.u.] and $l = 1$ [a.u.]. The fitting function is given by Eq. 6 under $C_I$ BC (instantaneous source).

By referring to Fig. 6 and the transport rate number magnitude obtained during the PADP data analysis, $N_R \approx 0.7$, it is possible to approximate the ratio between the apparent and actual diffusion coefficient for the problem under consideration, $D^{ap}/D^{ac} \approx 0.2$. Assuming the apparent diffusion coefficient to be equal $D^{ap} \equiv D_{C_I} = (1.35 \pm 0.03) \cdot 10^{-9}$ cm$^2$ s$^{-1}$ (exemplary value from the absorption profiles data analysis), and $D^{ac} \equiv D_{C_{III}} = (4.84 \pm 0.41) \cdot 10^{-9}$ cm$^2$ s$^{-1}$ (from PADP analysis), one obtains ratio $D^{ap}/D^{ac} = 0.29 \pm 0.03$, which is comparable to the value obtained from the simulations.

Summarizing, the mass absorption curves provide information on the overall transport kinetics by a certain system consisting of particular pair of donor and acceptor compartments (characterized by specific interfacial properties). It appears that the impact of the specific regions (bulks/interfaces) on the transport rate cannot be simply specified or quantified by means of a single absorption profile analysis. As such, results of the absorption curve analysis, yielding system characteristic values like $\tau$ or $D$ values for instance, should not be re-used for the direct analysis, modelling or accurate prediction of other systems properties.

## 4.5 EC-based analysis of PADP data

The PADP data analysis can be pursued in another direction, i.e. by adopting variable diffusion coefficient ($D(x)$) in the Fick's model, instead of the interlayer mass transfer hindrance effects analyzed by means of a certain form and behavior of boundary condition. The $D(x)$ is usually justified by a variable properties of an absorber system (i.e. spatial heterogeneity of the system). In case of DDC membrane synthesized here, there are, however, no clear premises to consider the absorber as heterogeneous. As such, the diffusion equation with $D(x)$ can be used as a parametrization tool allowing for the characterization of mass transfer rate variability across the membrane.

For the purpose of the PADP data analysis the EC-based approach is proposed. It is assumed, that the membrane is made up of layers, each characterized by resistivity and volume (in fact, the cross sections of layers, $a$, are equal, thus the volume is directly related to characteristic lengths of layers). In other words, a single system is divided into number of compartments. Mass transfer between the layers (compartments) can be further characterized by means of the Kirchhoff's laws (for more details see [22]). As an example, let us consider an absorber of thickness $l = 1$ [a.u.] and diffusivity $D = 1$ [a.u.], divided into two equivalent compartments of thickness $0.5l$ (i.e. the absorber domain $x \in (0; 1)$ is split into $x_1 \in (0; 0.5)$ and $x_2 \in (0.5; 1)$), in a one side contact with a donor ($x \leq 0$), and the donor is characterized by constant permeant concentration. As presented before, the mass absorption characteristics depend on the interfacial donor/acceptor behaviour ($C_I$ vs $C_{II}$ vs $C_{III}$ - see Fig. S1) and the transfer rate number magnitude (for high $N_R$, $C_{II}$ and $C_{III}$ mimics $C_I$ behaviour). Fig. 7a presents normalized absorption curves for the two acceptor compartments (black: $m(t)/m_{eq} = \int_0^{0.5} c(x,t)dx$, red: $m(t)/m_{eq} = \int_{0.5}^1 c(x,t)dx$, where $c(x,t)$ is given by Eqs 12 and $C_{III}$-type BC) for $N_R$=100, while inset demonstrates absorption curves for $N_R = 0.2$. For the $N_R$=100 case, the two curves follow different patterns and so represent distinct sorption characteristics despite similar intrinsic properties $R$ and $V$. As such, the electric circuit characteristic time, given simply by $\tau = \tau^n = RV$ is no longer valid, and no longer fulfils $m(t)/m_{eq} \approx 0.63$. For the $N_R = 0.2$ case, due to a slow interfacial equilibration, the $x_1 \in (0; 0.5)$ compartment is considered as a *resistive* layer, while the $x_2 \in (0.5; 1)$ layer almost instantaneously reaches the $m(t)/m_{eq}$ magnitudes of the first layer. Eventually, both profiles follow similar patterns, and reveal similar characteristic times. By following the EC-based mass transfer analysis presented in [22], an approximated value of characteristic time for *2nd* layer of a 2-layered system can be given as: $\tau = V_1 R_1 + V_2(R_1 + R_2) = \frac{l_1^2}{D_1} + l_1(\frac{l_1}{D_1} + \frac{l_2}{D_2})$ (subscripts denote subsequent layers of a system), and the absorption curve follows exponential character. As such, behaviour of systems displaying a certain mass hindrance effects of the first layer (or displaying interfacial barrier properties) can be easily predicted and modelled if thicknesses and diffusivities of the layers are known.

The PADP experimental data management for the EC-based analysis remain similar to the one described earlier; the scanning depth is related to the modulation frequency used (see Eq. 10), while the amplitude of PA signal ($S$) is proportional the total amount of pigment within a layer, $S \propto m$. The total amount of pigment within a certain ($i$-th) layer can be evaluated directly from the raw PA signals: $m_i(t) = S_{L_i}(t) - S_{L_{i-1}}(t)$, or from $c(x,t)$ data by subtracting signals from two neighbouring layers: $m_i(t) = \int_{L_{i-1}}^{L_i} c(x,t) dx$, where $L_i$ depends on the modulation frequency. Clearly, the total number of compartments depends on the number of modulation frequencies used, while the characteristic lengths of the compartments depend on the modulation frequency values picked for the experiment.

Absorption curves for a *layered* membrane, obtained from the PADP experimental data, are shown in Fig. 7b. The whole membrane was divided into two regions: one directly adjacent to the drug donor and of thickness of around 4 µm, and the second one encompassing rest of the absorber and of thickness of ~18 µm. It can be noted immediately, that the absorption curves follows the same pattern, which implies the first layer acts as *the resistive* one. Quantitative interpretation of the data can be performed in terms of Kirchhoff's current law-based analysis, yielding set of two ODE's:

$$\dot{m}_I(t) = l_I \dot{c}_I(t) = \frac{c_0 - c_I(t)}{R_I} - \frac{c_I(t) - c_{II}(t)}{R_{II}}, \tag{Eq. 13a}$$

$$\dot{m}_{II}(t) = l_{II} \dot{c}_{II}(t) = \frac{c_I(t) - c_{II}(t)}{R_{II}}, \tag{Eq. 13b}$$

with the initial conditions:

$$c_0 = 1, c_I(0) = c_{II}(0) = 0, \tag{Eq. 13c}$$

where subscripts indicate specific compartment ("0" – drug donor, "*I*" – membrane compartment ($l_I$ assumed to be equal to 4 µm) adjacent to the donor, "*II*" – the outer membrane compartment of thickness $l_{II} = 18$ µm). The best-fit curves of the model given by Eqs. 13 to experimental data are represented by solid lines in Fig. 7b. The best-fit parameters of the model were found to be: $R_I = 77.2 \pm 1.9$ µm⁻¹s and $R_{II} = 3.62 \pm 2.31$ µm⁻¹s. By taking simple $\tau^n = RV = Rl$, the layer resistances can be translated to $\tau_I^n = 308 \pm 7.6$ s and $\tau_{II}^n = 65.2 \pm 41.6$ s; then, by assuming $N_R$ stands for a ratio of characteristic times representing bulk and interfacial processes, $N_R \approx \tau_{II}^n/\tau_I^n \approx 0.2$ (in fact this is a certain simplification of the problem, as the $\tau_I^n$ parameter involves contributions from both interfacial and bulk transfers). It can be noted, that the value is in the same order of magnitude as the transport rate number obtained during the distributed PADP data-set analysis under $C_{III}$ boundary condition.

As the first membrane layer displays barrier properties, the mass uptake can be modelled by means of single exponential model. By taking resistances and characteristic lengths of the layers as obtained before, the predicted characteristic time of the system is $\tau_p = l_I R_I + l_{II}(R_I + R_{II}) = 0.49 \pm 0.02$ h, and is comparable to the exponential fitting parameter $\tau_{bf} = 0.45 \pm 0.02$ (simple monoexponential best-fit to data for $x_I \in (0$ µm; $4$ µm$)$ from Fig. 7b, not shown). The both EC-based characteristic times remain in a considerable agreement with the values obtained earlier from the direct PA data analysis (profilometric data analysis under the constant source condition: $\tau_{C_I}^{EC} = 0.45 \pm 0.01$ h, spectroscopic data analysis: $\tau_{PAS}^{EC} = 0.42 \pm 0.02$ h).

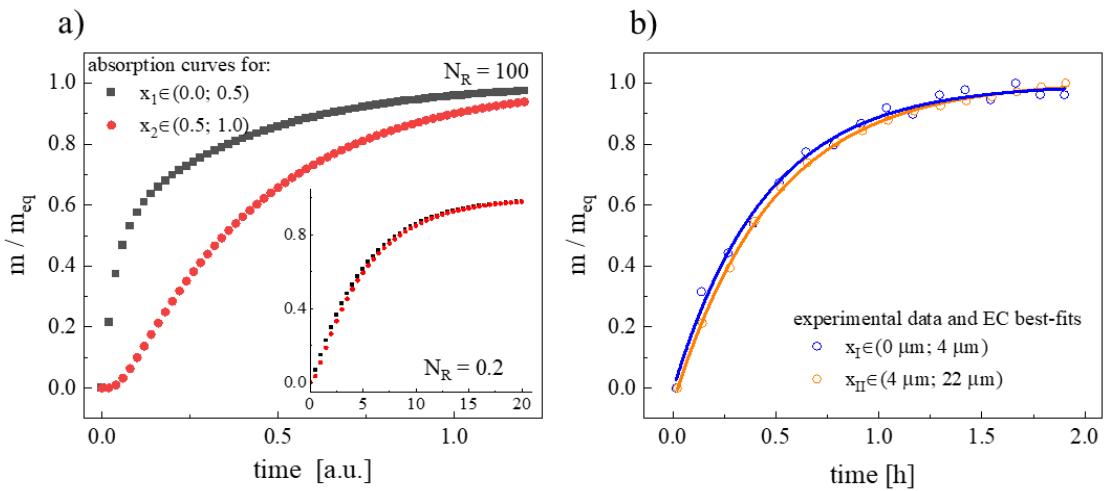

Fig. 7 a) normalized absorption (diffusion model under $C_{III}$-type BC) curves for two adjoined acceptor compartments ($D = 1$, black: $x_1 \in (0; 0.5)$, red: $x_2 \in (0; 1)$) in contact with donor compartment ($x < 0$) for $N_R$=100 (main) and $N_R = 0.2$ (inset); b) experimental data for a *layered* membrane (first layer of ~4 µm, second layer ~18 µm) and the best-fits under the EC-based model given by Eqs 13.

## 5. Conclusions

The research focuses on some issues related to absorption/release data analyses in reference to actual processes influencing mass transport dynamics. In particular it is argued whether the sorption curves, delivered experimentally and validated theoretically by means of various approaches to mass transport phenomena, can reflect physics underlying drug absorption (release), or allow for a reliable identification and quantification of the transport parameters.

The mass transport models investigated were based on the Fick's diffusion model under three types of boundary conditions: $C_I$ – the Dirichlet constant source (standard approach), $C_{II}$ – the Neumann's interfacial concentration gradient condition and $C_{III}$ – the Dirichlet's interfacial equilibration condition. It appeared, that despite noticeable differences in the predicted concentration field evolutions ($C_I$ *vs* $C_{II}$ and $C_{III}$), all the models provided similar absorption curves (the total amount of mass absorbed *vs* time) in terms of their curvatures. The results allow to question the reliability of the absorption curves for the transport parameter determination.

The $C_I$ and $C_{III}$ models were confronted against experimental data acquired for a transdermal delivery surrogate system by means of two photoacoustic modalities, spectroscopy (PAS) and depth-profiling (PADP). PAS allowed to track the total pigment amount inside the membrane, as well as monitor stability of the drug, while PADP provided time-dependent drug distribution inside the acceptor system. The PADP profilometric (full) dataset evidenced mass distribution behaviour (especially close to the donor/acceptor interface) which cannot be easily interpreted by the instantaneous source hypothesis. In fact, the data analysis performed by means of the $C_{III}$–based approach suggested that the overall mass uptake is almost equally influenced by interfacial and bulk transport processes. The analyses were performed by means of newly introduced parameter, the transport rate number, given by a ratio of characteristic times of bulk diffusion and interfacial equilibration ($N_R = \tau_D^n / \tau_{eq}$).

The analysis of the PAS and PADP (reduced dataset) absorption profiles yielded comparable results by means of diffusion coefficient and characteristic times. In fact, the curvature of both datasets remained very similar. Despite very convincing fitting quality (by means of the *1st* order model), the diffusion coefficients (and so characteristic times) obtained were significantly different from the diffusion coefficient obtained during PADP profilometric data analysis under $C_{III}$ BC. It appeared, that various approaches (transport models under certain BCs) may lead to similar predictions of the $m(t)$ in terms of the function curvature. This makes the $m(t)$ data of limited reliability for the absorption/release mechanisms identification and quantification.

The last step of the research was devoted to the data analysis by means of the EC model. The EC approach allowed for the determination of characteristic times of transport processes for a *layered* membrane, where the first layer was considered to be directly adjacend to the drug donor phase (as such, the characteristic time of the layer was related to the characteristic time of the interfacial processes), while the second layer represented the bulk transport. The transport rate number obtained was in a considerable agreement with the value obtained during the PADP data analysis. The EC-based analysis of absorption curves provided characteristic time of the transport process comparable to the values obtained during PAS and PADP analysis by means of standard approaches.

The paper underlines the need for further depth-profiling studies on the mass transfer in viscous systems. It is expected, that such studies will allow for a better understanding of the foundational underpinnings of the membrane transport, which may trigger further studies on a novel delivery systems development strategies.

**Supplementary materials**

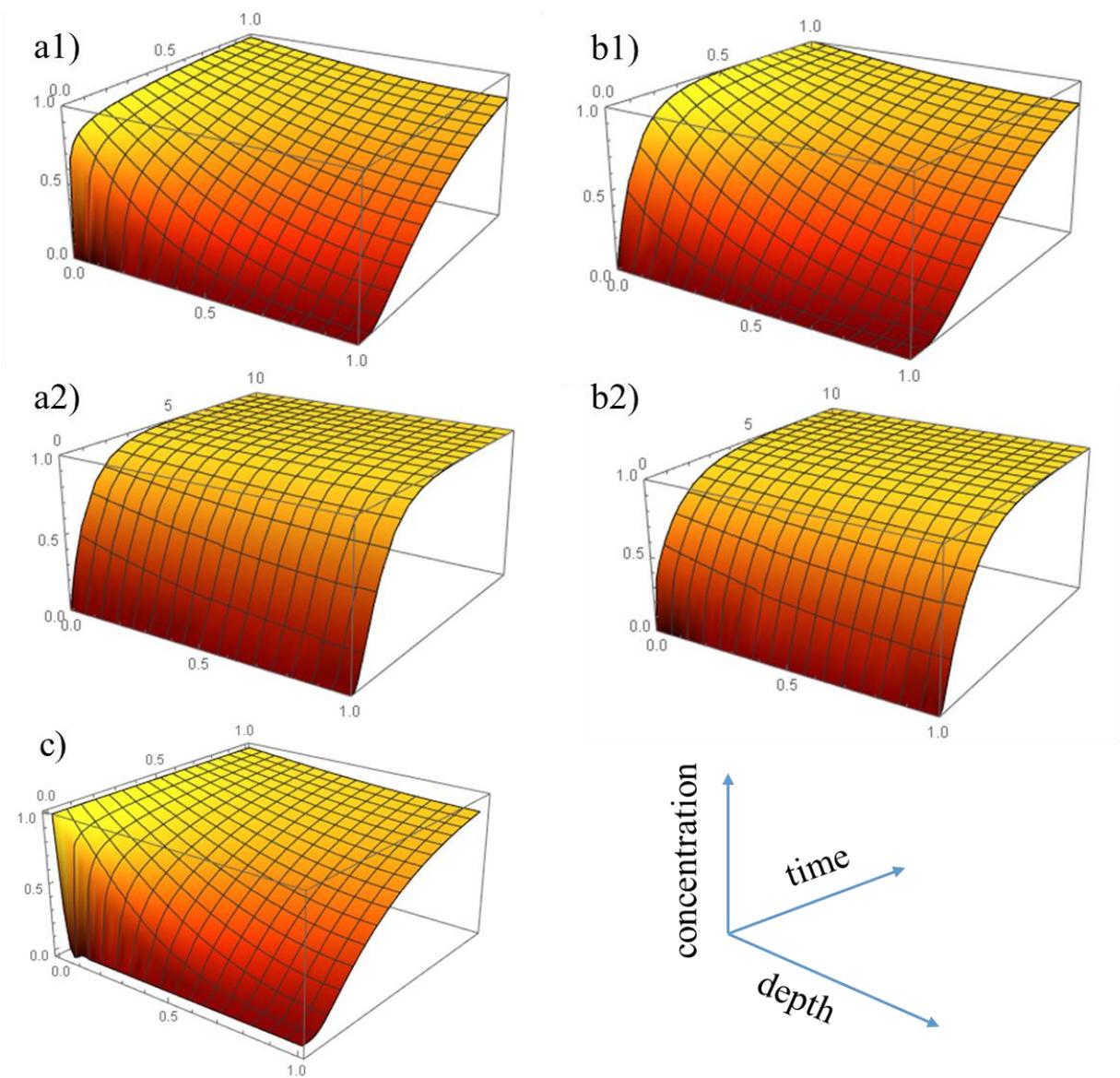

Fig. S1 The evolution of the diffusion-controlled (the acceptor is characterized by $D = 1$, $l = 1$) concentration profiles ($c(x,t)$ data) under $C_{II}$ (a) and $C_{III}$ (b) BC for relatively low $N_R$s (upper row: $N_R = 10$, middle row: $N_R = 1$). The evolution of the concentration field under $C_I$ (constant source) is shown in (c).

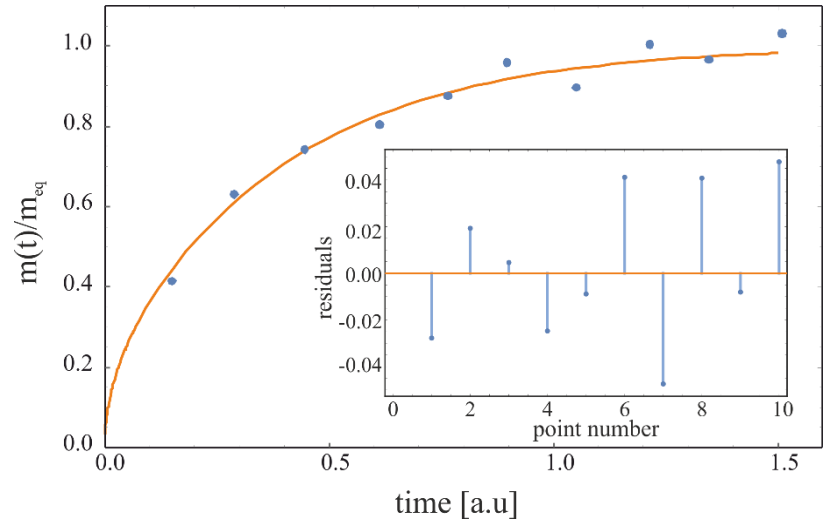

Fig. S2 Exemplary artificial absorption profile (the upper limit of measurement precision of 5%, 10 experimental points distributed within $t\epsilon(0; 5\tau)$) obtained by means of diffusion-driven transport model (Eqs 12) for a finite membrane of diffusivity $D = 1$ [a.u.] and thickness $l = 1$ [a.u.], under constant concentration at the source boundary condition ($C_I$ condition).

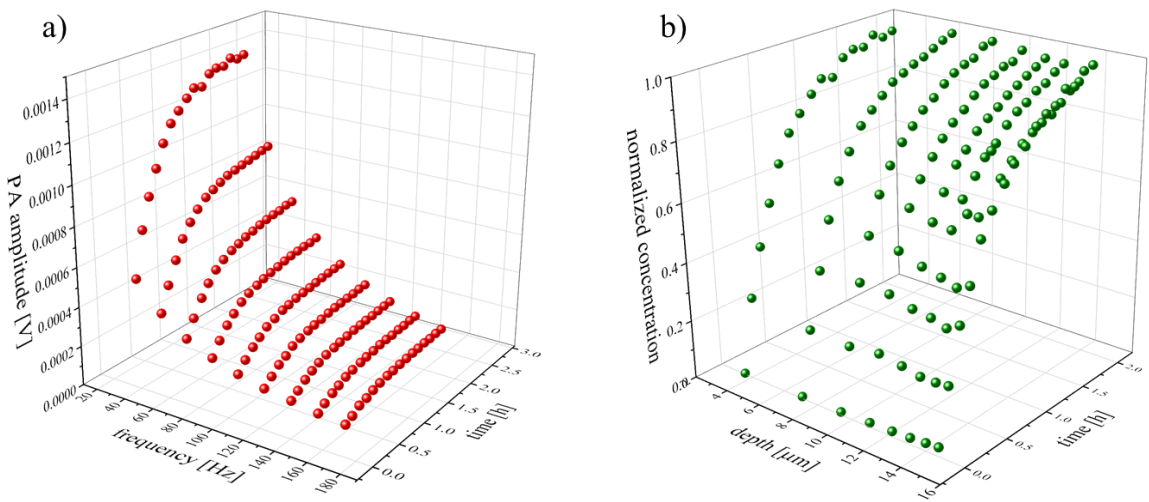

Fig. S3 Results of the photoacoustic depth-profiling studies on the membrane transport. Raw data points are shown in (a), while the time-dependent drug concentration profiles are shown in (b).

Tab.S1 Best-fit diffusion coefficient (mean and standard deviation) of various absorption models to artificial 1000 data-sets generated by means of the diffusion driven transport model (Eqs 12 under $C_I$ BC), characterized by distinct number of experimental points (3, 5, 10 or 20 points) *measured* with a certain *precision* upper limit (1%, 3% or 5%) and randomly distributed within a certain time interval.

| model | precision [%] | *Diffusion coefficient, mean (standard deviation),* long experimental window $t\epsilon(0; 5\tau)$ | | | |
|---|---|---|---|---|---|
| | | number of points | | | |
| | | 3 | 5 | 10 | 20 |
| exact | 1 | 1.00 (0.03) | 1.00 (0.02) | 1.00 (0.03) | 1.00 (0.03) |
| | 3 | 1.01 (0.08) | 1.00 (0.07) | 1.00 (0.08) | 1.01 (0.08) |
| | 5 | 1.01 (0.13) | 1.01 (0.13) | 1.01 (0.13) | 1.01 (0.13) |
| EC mono-exp | 1 | 0.98 (0.03) | 0.99 (0.02) | 1.00 (0.03) | 1.00 (0.03) |
| | 3 | 0.99 (0.07) | 0.99 (0.07) | 1.00 (0.07) | 1.01 (0.07) |
| | 5 | 0.99 (0.12) | 0.99 (0.12) | 1.01 (0.13) | 1.01 (0.13) |
| Higuchi | 1 | 0.62 (0.02) | 0.65 (0.01) | 0.66 (0.01) | 0.68 (0.01) |
| | 3 | 0.62 (0.03) | 0.65 (0.03) | 0.66 (0.03) | 0.67 (0.03) |
| | 5 | 0.62 (0.05) | 0.65 (0.04) | 0.66 (0.04) | 0.68 (0.05) |

| model | precision [%] | *Diffusion coefficient, mean (standard deviation),* short experimental window $t\epsilon(0; \tau)$ | | | |
|---|---|---|---|---|---|
| | | number of points | | | |
| | | 3 | 5 | 10 | 20 |
| exact | 1 | 1.00 (0.02) | 1.00 (0.02) | 1.00 (0.03) | 1.00 (0.03) |
| | 3 | 1.00 (0.07) | 1.00 (0.07) | 1.00 (0.07) | 1.00 (0.08) |
| | 5 | 1.00 (0.12) | 1.00 (0.12) | 1.00 (0.13) | 1.00 (0.13) |
| EC mono-exp | 1 | 1.04 (0.02) | 1.05 (0.02) | 1.05 (0.02) | 1.05 (0.02) |
| | 3 | 1.05 (0.07) | 1.05 (0.07) | 1.05 (0.07) | 1.05 (0.07) |
| | 5 | 1.04 (0.11) | 1.05 (0.12) | 1.05 (0.12) | 1.05 (0.12) |
| Higuchi | 1 | 0.99 (0.02) | 0.99 (0.02) | 0.99 (0.02) | 0.99 (0.02) |
| | 3 | 0.99 (0.07) | 0.99 (0.07) | 0.99 (0.07) | 0.99 (0.08) |
| | 5 | 0.99 (0.11) | 0.99 (0.12) | 0.99 (0.12) | 0.99 (0.13) |